\documentclass{WileyMSP-template}

\begin{document}
\RaggedRight
\pagestyle{fancy}

\title{Optically manipulated micromirrors for precise\\excitation of WGM microlasers}

\maketitle


\author{Tomasz Plaskocinski}
\author{Libin Yan}
\author{Marcel Schubert}
\author{Malte C. Gather}
\author{Andrea Di Falco*}


\dedication{}

\begin{affiliations}

Mr.\ Tomasz Plaskocinski\\
School of Physics and Astronomy, University of St Andrews, North Haugh, St Andrews, KY16 9SS,\\ Scotland, UK\\
Email Address: tp43@st-andrews.ac.uk\\
\medskip
Dr.\ Libin Yan\\
School of Physics and Astronomy, University of St Andrews, North Haugh, St Andrews, KY16 9SS,\\ Scotland, UK\\
Email Address: sglbyan@gmail.com\\
\medskip
Dr.\ Marcel Schubert\\
Humboldt Centre for Nano- and Biophotonics, Department of Chemistry, University of Cologne,\\ Greinstr. 4--6, 50939 Cologne, Germany
Email Address: marcel.schubert@uni-koeln.de\\
\medskip
Prof.\ Malte C. Gather\\
(1) School of Physics and Astronomy, University of St Andrews, North Haugh, St Andrews, KY16 9SS,\\ Scotland, UK\\ \medskip
(2) Humboldt Centre for Nano- and Biophotonics, Department of Chemistry, University of Cologne,\\ Greinstr. 4--6, 50939 Cologne, Germany\\
Email Address: mcg6@st-andrews.ac.uk\\
\medskip
Prof.\ Andrea Di Falco\\
School of Physics and Astronomy, University of St Andrews, North Haugh, St Andrews, KY16 9SS,\\ Scotland, UK\\
Email Address: adf10@st-andrews.ac.uk

\end{affiliations}




\begin{abstract}

Whispering gallery mode microlasers are highly sensitive refractive index sensors widely explored for biophotonic and biomedical applications. Microlaser excitation and collection of the emitted light typically utilize microscope objectives at normal incidence, limiting the choice of the oscillation plane of the modes. Here, we present a platform that enables the excitation of microlasers from various directions using an optically manipulated micromirror. The scheme enables precise sensing of the environment surrounding the microlasers along different well-controlled planes. We further demonstrate the capability of the platform to perform a time-resolved experiment of dynamic sensing using a polystyrene probe bead orbiting the microlaser.

\end{abstract}


\section{Introduction}
Non-invasive interactions such as sensing, illumination, or touching of biological samples are essential for understanding their properties and behaviors. Optical microscope objectives are the primary tool for these optical investigations. However, objective-based methods only offer keyhole-like access to the sample, resulting in limited light delivery and collection angles. Objectives are also bulky and operate relatively far from the observation plane, further limiting the interaction with the biological object of interest. 

In recent years, lab-on-chip platforms have revolutionized the field of biophotonics, giving access to specimens with many degrees of freedom, e.g.\ for applications in high throughput sensing, imaging and multi-technique processing of analytes~\cite{bios12070501, Huang2015,Manefjord2022,ZhangWongZengBiTaiDholakiaOlivo+2021+259+293}. A particularly useful paradigm is the use of microresonators, which enhance the light-matter interaction, through increased contact time with the analyte through, e.g., photonic crystal sensing\cite{Pitruzzello_2018,DiFalco2009} and ring resonators\cite{nitkowski2011chip,Steglich2017HybridWaveguideRR}. 

Whispering Gallery Mode (WGM) resonators are amongst the most versatile types of microcavities and come in a variety of different geometries, such as spheres\cite{Schubert2015LasingWL}, disks\cite{Martino2019WavelengthencodedLP,Fikouras2018NonobstructiveIN}, and toroids\cite{Polman2004UltralowthresholdET}, both in 
solid\cite{Steglich2017HybridWaveguideRR,Martino2019WavelengthencodedLP,Fikouras2018NonobstructiveIN} and liquid materials\cite{Humar2015IntracellularM,Mcgloin2017DropletLA}. They support optical modes which strongly depend on the resonators' size, refractive index, and shape but are also highly sensitive to the refractive index in the close proximity 
($\sim$100 nm) of the resonator surface. This makes them excellent candidates for refractive index sensing\cite{https://doi.org/10.1002/adom.202300530}, single molecule detection\cite{Yu2021}, optomechanics\cite{Kippenberg:07}
and metrology\cite{doi:10.1126/science.1193968} applications. 

WGM resonators complement other fluorescence-based imaging and sensing techniques\cite{Wang2022DemonstrationOI,Toropov2021ReviewOB}. Due to the high Q-factor of WGM resonators, the combination with gain media enables efficient laser-based sensing. Due to their small footprint, these microlasers can be embedded into living biological samples, allowing for various new interrogation methods, such as simultaneously 
barcoding and tracking thousands of cells\cite{Schubert2015LasingWL,Humar2015IntracellularM,Martino2019WavelengthencodedLP}. WGM lasers have also been used to measure intracellular forces: for example, dye-doped polystyrene microlasers embedded in mouse heart cells were used to detect precise contractility profiles of single cells by monitoring the shift in the refractive index around the microlaser\cite{Schubert2020}. 
Recent progress has also shown the first \textit{in-vivo} applications of microlaser sensors\cite{Schubert2020,Li2020OpticalCT,Li2022InVT}, demonstrating the growing complexity of the biological environments that can be investigated with these techniques.

Microlasers are typically excited by pumping the sample at normal incidence using microscope objectives and either small, focused beams or wide-field excitation. The pumping direction and geometry are crucial 
for the laser operation as they control the plane of the laser oscillation and, therefore, the volume that can be probed by the evanescent field of the WGM.\ Changing the geometry of the spherical microlasers 
can lift the laser mode degeneracies and permit to discriminate between oscillating planes at the cost of reduced control of the lasing modes\cite{Weiss:95}.
Furthermore, there have been efforts to try and 
decouple the pumping and detection directions\cite{Rafferty2019OpticalDO}. Most of the current issues are related to the use of a single objective for the excitation and detection of the microlasers, which provides little control over the exact lasing and collection planes. These constraints hamper further applications of microlasers, especially in complex cellular or tissue environments.

Here, we present a platform that extends the versatility of microscopic optical sensors by introducing various degrees of freedom (angle, distance, direction) for the excitation of WGM microlasers formed by dye-doped polystyrene microspheres. To this end, we fabricate polymeric membranes, decorate them with handles for optical manipulation, and use them as \textit{in-situ} movable mirrors. This approach overcomes the typical keyhole 
limitation of objectives, as it is possible to freely select the lasing plane in spherical microlasers. We demonstrate that the wavelength of the lasers shifts as the mode interacts 
with environments of different refractive indices. We thoroughly calibrate the system using microlasers suspended in water and placed on glass and show that the micromirrors can excite WGMs 
along different axes of the microlasers, as shown in \textbf{Figure 1} below. Finally, we perform a proof-of-concept experiment where we orbit a microlaser with a polystyrene probe bead and detect the shift in the lasing wavelength 
brought about by the interaction between both objects. 

\begin{figure}[ht!]
\centering\includegraphics[width=16cm]{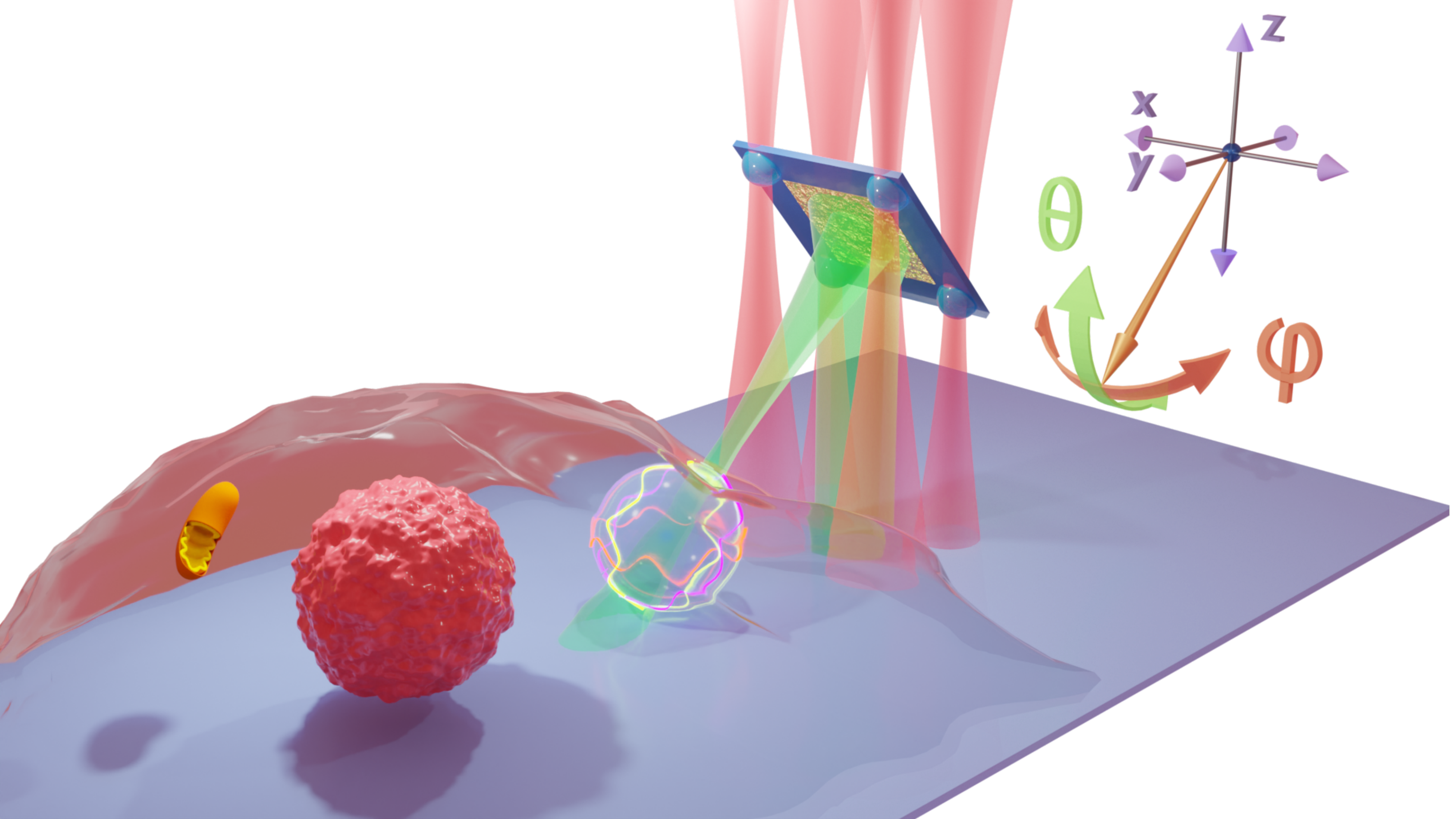}
\caption{Scheme using an optically trapped mirror to excite WGMs across different planes of a spherical microlaser suspended on glass. Also shown is the coordinate system used throughout the paper when discussing the micromirror tilt. Before incidence on the
micromirror, the trapping and excitation laser beams propagate along the positive z direction.  The center of the micromirror is chosen as a reference point.}\label{fig:fig1}
\end{figure}

\section{Materials and Methods}
\subsection{Micromirror Fabrication}
Details of the fabrication are shown in \textbf{Figure 2}. The design of the polymeric membrane with a gold film was optimized for controlling the pumping of spherical microlasers, with the parameters shown in Figure 2a. The membrane width is $w_m$ = 15 \textmu m, the thickness is $t_m$ = 200 nm, the width of the gold mirror is $w_g$ = 8 \textmu m, the gold thickness is $t_g$ = 35 nm, the diameter of the handles is $d$ = 2 \textmu m, and the height of the handles is $h$ = 1.5 \textmu m. A Scanning Electron Microscope (SEM) image of a finished micromirror is shown in Figure 2b.

The SU8 membrane width $w_m$ was chosen to allow for sufficient tilting, considering the depth of focus of the trapping objective (see details below). The gold patch was sufficiently large to focus the pump beam without it interfering with the optical 
trapping of the micromirror. The handle size was chosen to enable stable trapping at the wavelength of 830 nm\cite{Askari:21}. The membrane thickness was chosen to maintain rigidity while managing the total mass of the micromirror for ease of manipulation. 

\begin{figure}[ht!]
\centering\includegraphics[width=16cm]{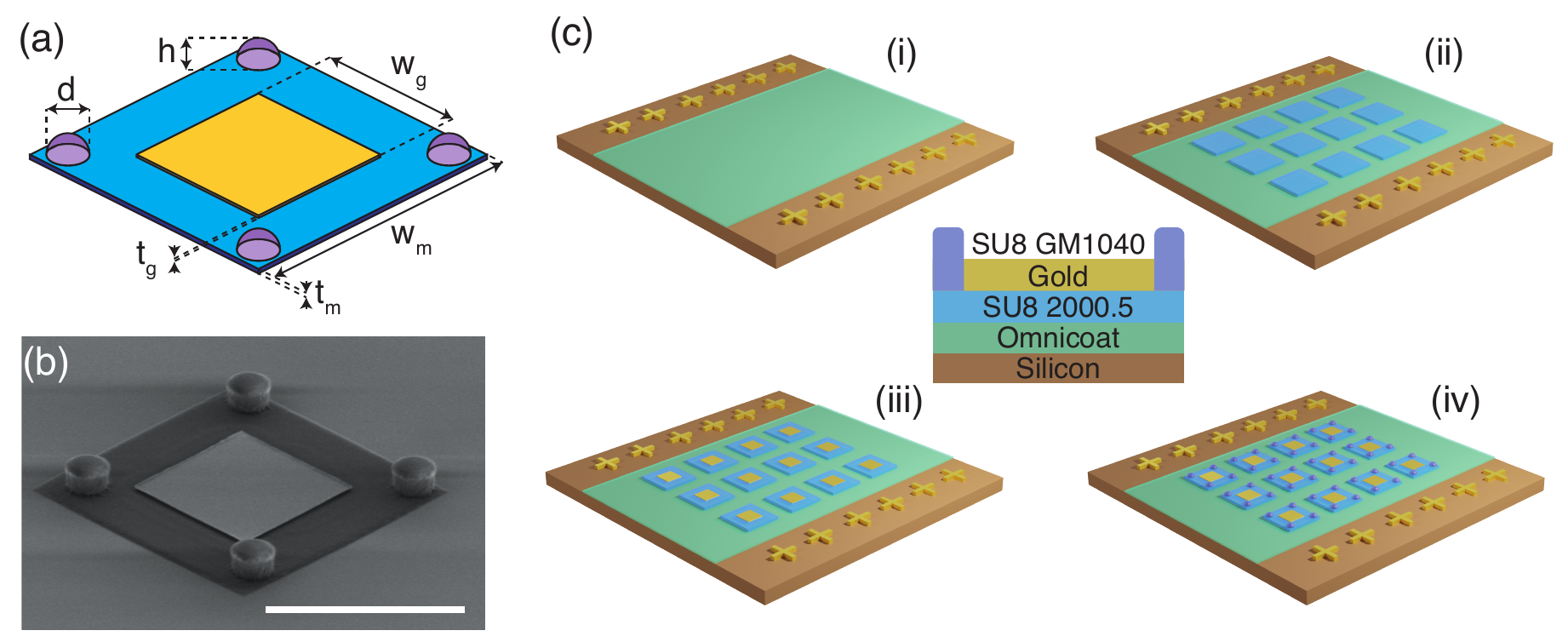}
\caption{a) Geometrical parameters of the micromirror: $h$ and $d$ are the handle height and diameter respectively, $w_g$ is the width of the gold film square, $t_g$ is the gold thickness, $w_m$ is the SU8 membrane width, and $t_m$ is the SU8 membrane thickness. b) SEM image of a micromirror. The scale bar is 10 \textmu m. The image of an array of micromirrors is shown in Figure S1, Supplementary Information. c) Key fabrication stpdf. Features size is exaggerated to present them clearly. 
(i) Gold alignment markers are written and deposited, and a sacrificial layer of Omnicoat is spun on a silicon carrier. (ii) SU8 membranes are written and developed. (iii) The gold film is deposited. (iv) SU8 handles are written and developed.}
\label{fig:fig2}
\end{figure}

The micromirror fabrication process is shown in Figure 2c, and completed using an Electron Beam Lithography (EBL) system (Raith e-Line Plus). We first defined alignment markers using a standard lift off process based on Polymethyl methacrylate (PMMA). The markers were made of 35 nm of Au on a 5 nm NiCr layer for improved silicon substrate adhesion. We then deposited a sacrificial layer of Omnicoat, followed by the SU8 film, defined the membranes shape with an EBL exposure, and developed the sample. The process using PMMA and metal evaporation was repeated, depositing the gold used as a mirror on the SU8 membrane. SU8 was then spin-coated, exposed with the EBL and developed, creating the handles for optical manipulation.

Finally, the sample was cleaved to facilitate access to the sacrificial layer, and dissolved with Tetramethylammonium Hydroxide (TMAH), lifting off the micromirrors, which were then exchanged into water. 

\begin{figure}[ht!]
\centering\includegraphics[width=16cm]{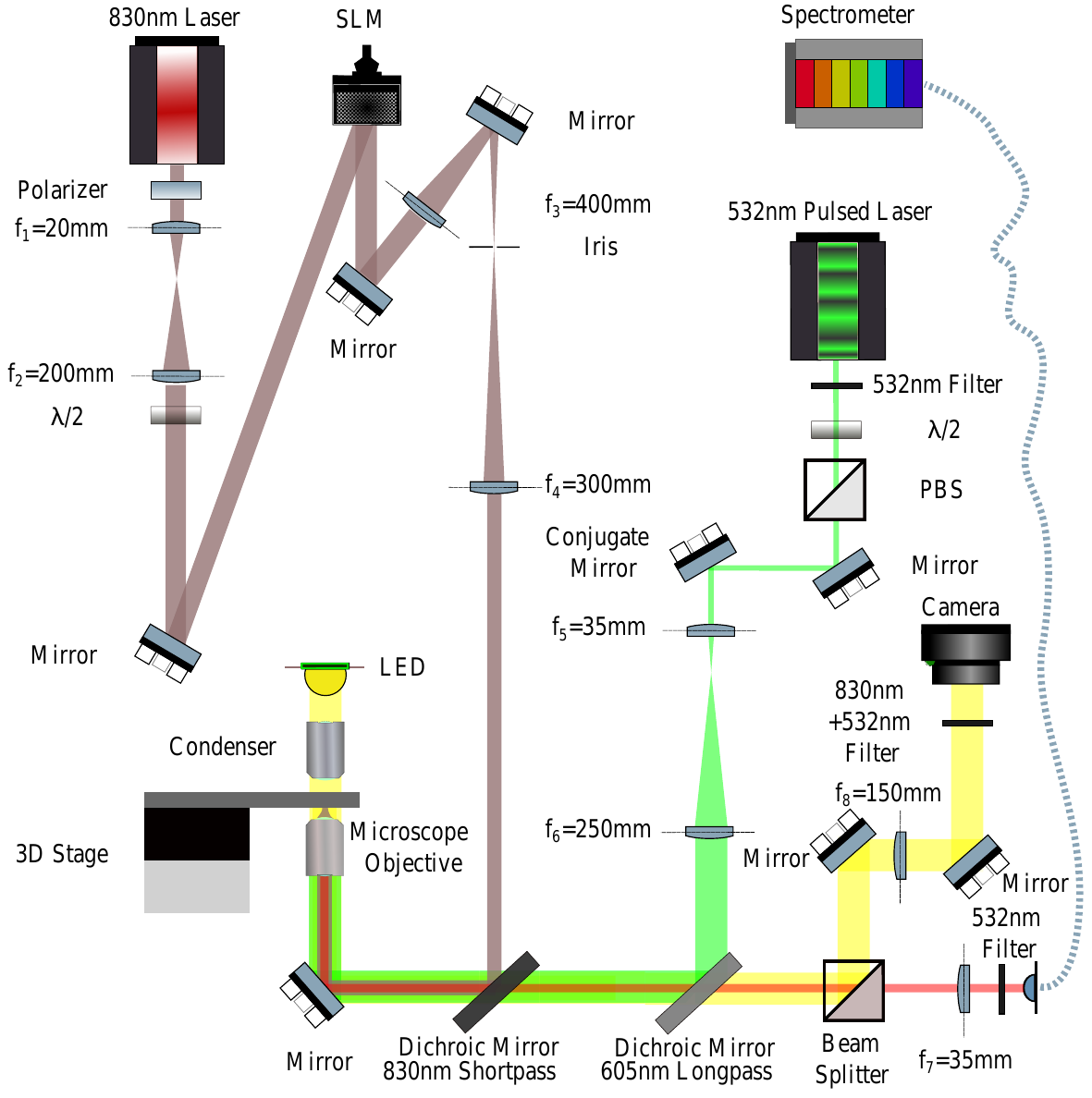}
\caption{Optical setup used for micromirror manipulation (brown, left), fluorescence excitation (green, center), fluorescence emission (red) and imaging (yellow, right). The focal lengths of lenses are denoted as $f_1-f_8$, the half-wave plate as $\frac{\lambda}{2}$, the spatial light modulator as SLM, anFd polarizing beam splitter as PBS.}
\end{figure}
\label{fig:fig3}
\subsection{Optical Setup}

As shown in \textbf{Figure 3}, the setup consisted of holographic optical tweezers, a pump and probe system, and an imaging system. Unless stated otherwise, all optical components came from Thorlabs.
The optical tweezers used a 230 mW continuous wave laser (LuxX\textsuperscript{\tiny\textregistered} 830-230, Omicron) with a wavelength $\lambda_t$ = 830 nm. The beam polarization was rotated to match the reflective Spatial Light Modulator (SLM) polarization axis (E-Series 1920 x 1200, Meadowlark). The beam was then sent through a 4f system consisting of lenses 3 and 4 and an iris placed in order to reject unwanted diffraction orders from the SLM. This 4f system imaged the SLM plane onto the back focal plane of the water immersion objective, overfilling it in order to use its full numerical aperture (UPLSAPO60XW, NA=1.2, Olympus).

The pump line consisted of a $\lambda_p$ = 532 nm laser with a pulse duration $\leq$1.3 ns (FDSS 532-Q3, CryLas), operating at a frequency of 100 Hz. The pump power was controlled using a combination of a half-wave plate and a polarizing beam splitter (PBS). A conjugate mirror was placed before a 4f system formed by lenses 5 and 6 for beam steering. Translating lens 6 along the optical
axis allowed adjustment of the focal plane of the pump laser relative to the imaging plane of
the objective.

The emission and imaging collection paths were separated using a 50:50 non-polarising beamsplitter. The microlaser emission collected by the microscope objective was sent to the spectrometer (328i, Kymera), operating with an integration time of 0.1 s, through a multimode fiber (400 - 700 nm AR coated, 105 \textmu m core diameter). The grating used was 1200 l/mm with a 532 nm blaze. The slit was 10 \textmu m wide. The spectroscopic camera (Newton, 970 EMCCD, Andor)  acquired the spectra at 100 Hz, with no Electron Multiplier Gain, and at -70° C. The resolution of the spectrometer was 35 pm.

The imaging system comprised a white LED, a condenser, and a camera (piA640-210gm, Basler). The LED was turned off for pump and probe experiments. Further information regarding the optical trapping setup are provided in Section 5. 

\section{Results and Discussion}
\subsection{Systematic excitation of WGM on different planes}
Probing the non-uniform environment surrounding the WGM microlasers formed by 15 \textmu m
dye-doped polystyrene microspheres (see Section 5 for additional details) at different angles presents a formidable challenge that is hard to achieve with a single objective. Shown in \textbf{Figure 4} is our calibration process. We took two distinct spectra taken by directly exciting the microlaser using the microscope objective, as shown in panels (i) and (ii) of Figure 4a. Due to the used pump geometry, the WGMs directly excited by the microscope objective circulate in the vertical plane and, thus, interact with the ambient medium located just below the microlaser. The spectrum (i) served as a reference for the condition of the microlaser 
sitting on glass (with refractive index $n_g\sim$1.52); the spectrum (ii) served as a reference for the condition of the microlaser surrounded only by water (with refractive index $n_w\sim$1.33). Figure 4b shows a typical set of the six WGM modes for the (i) case, which were fitted to extract their width, amplitude, and central wavelength using a Gaussian fitting. 

\begin{figure}[ht!]
\centering\includegraphics[width=16cm]{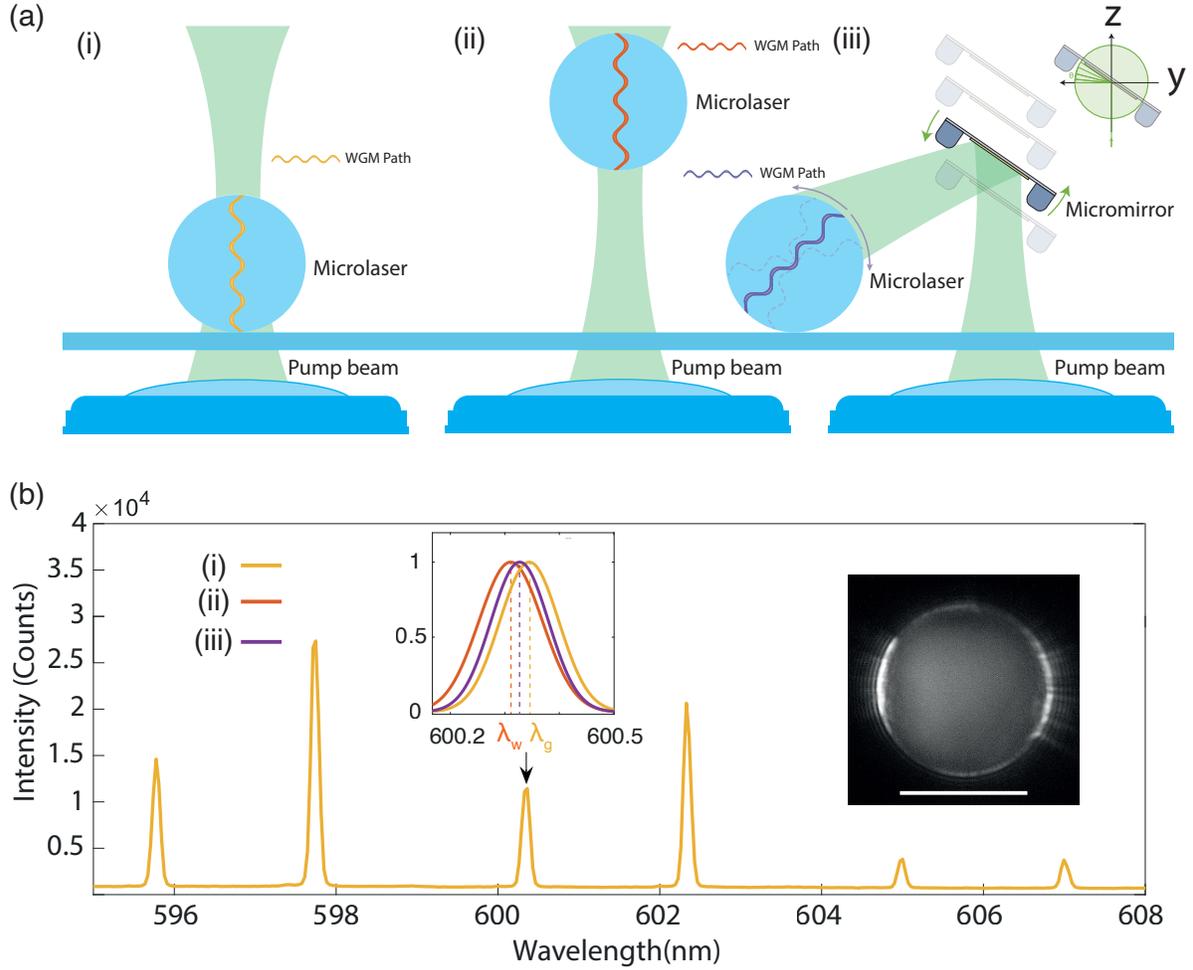}
\caption{a) Sketch of the illumination conditions of the WGM microlaser while in contact with (i) glass (using objective), (ii) water (using objective), and (iii) glass (using micromirror). The colored wavy lines indicate the paths of excited WGMs circulating in the microlaser. b) Typical WGM
lasing spectrum for case (i). The inset on the left shows the fitted normalized spectra of the mode marked by the arrow for all cases (i)-
(iii), with line colors matching the colors of the corresponding WGM paths shown in part (a). The mode (iii) excited by the micromirror is shifted between the water (ii) and glass (i) modes. The inset on the right shows a camera image of an illuminated microlaser in the case of (i). The scale bar is 10 \textmu m.}
\label{fig:fig4}
\end{figure}

As illustrated in Figure 4a(iii), the trapped micromirror was then brought in and positioned using the SLM. The micromirror illuminated the microlaser at various heights and tilt angles $\theta$ around the x-axis, and $\varphi$ around the z-axis, as shown in Figure 1, resulting in the excitation of WGMs circulating in oblique planes. The left inset of Figure 4b shows the normalized spectrum of one of the modes for the three cases (i)-(iii).
The full mapping of the illumination positions can be found in Figures S4 and S5, Supplementary Information. 
The mode spectral shifts observed for
five different heights of the micromirror, averaged over the six WGMs shown in Fig. 4b, are shown in \textbf{Figure 5a}, each for 20 combinations of angles scanned in $\theta$ and $\varphi$. 

Five sets of 20 spectra were taken: four sets for the microlaser sitting on glass [cases (i) - (iv)] and one where the microlaser was entirely suspended in water [case (v)], for which we obtained the parameters of the modes at specific $z$, $\theta$, and $\varphi$. 
The spectra were processed by first taking away the corresponding background to offset slight variations of the collection plane of the microscope for different positions of the micromirror.

In Figure 5a, the spectral shifts are referenced to the condition of the microlaser in contact with the glass ($\Delta\lambda_g$). For clarity, we also mark the shift of the spectra with respect to the reference 
($\Delta\lambda_w$), when the microlaser was trapped and fully suspended in water. For both reference spectra, the microlaser was pumped with the objective in the orthogonal orientation without using the micromirror [see Figure 4a, panels (i) and (ii)]. The positions of the mirror and microlaser are shown in Figure 5b, denoted with labels (i)-(v) for different heights of the micromirror above the glass slide plane. The vertical axis in Figure 5a shows the shifts for four different 
angles $\theta$ ranging between 25\degree and 40\degree, each probed for five angles $\varphi$ ranging between -10\degree and 10\degree. The radius of each data point is dictated by the sum of the peak values of the six most intense WGMs in each spectrum (see e.g. Figure 4b). In each panel (i-v), the radii of data points are normalized to the highest value for the specific height of the micromirror. This helps to inform which spectra are the strongest and, therefore, most likely to be accurate, due to a high signal-to-noise ratio (SNR). In each panel of Figure 5a, the dashed lines indicate the mean spectral shift of the peaks averaged over all used illumination angles $\theta$ and $\varphi$. The dotted lines indicate the standard deviation of the shifts.  
Additionally, on the left of Figure 5a, we plotted histograms of all the shifts, which show how the data points cluster around the spectral shifts $\Delta\lambda_w$ and  $\Delta\lambda_g$. Two dimensional representations of the averaged spectral shifts as functions of $\theta$ and $\varphi$ are shown as heatmaps in Figure S6, Supplementary Information. 

\begin{figure}[ht!]
\centering\includegraphics[width=16cm]{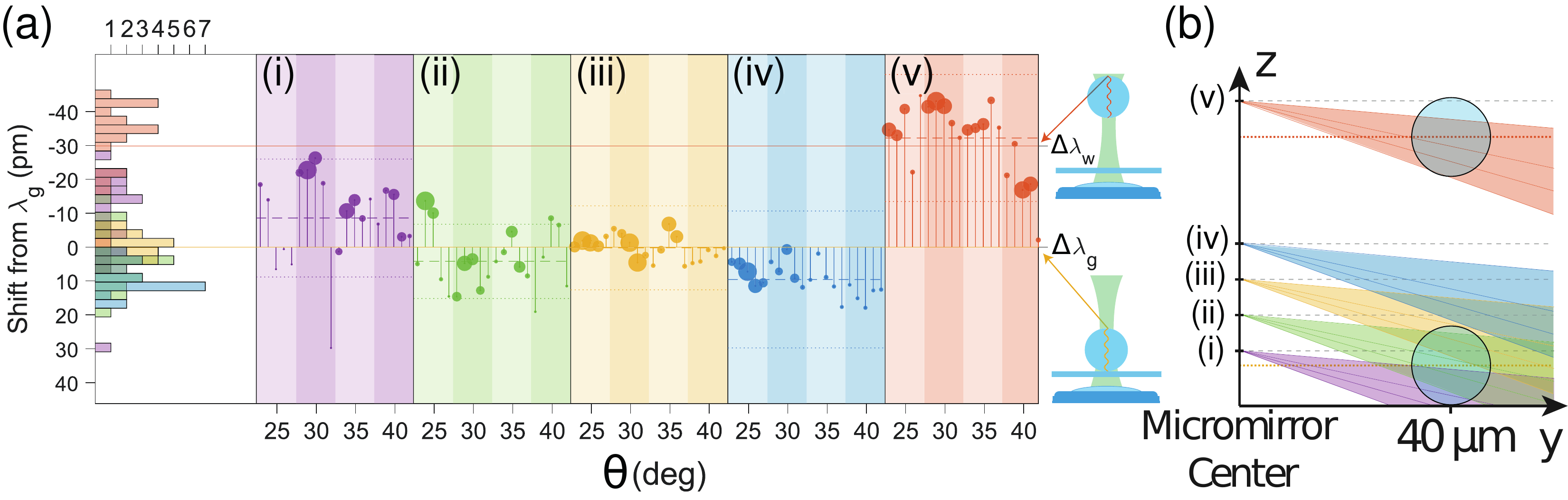}
\caption{a) Average shift of spectra excited using the micromirror placed at different heights (i) 10 \textmu m (ii) 17 \textmu m (iii) 23 \textmu m (iv) 30 \textmu m and (v) 57 \textmu m above the glass slide plane. The shift is relative to the spectrum of the microlaser sitting on glass $\lambda_g$, acquired using direct excitation through the microscope objective. The size of points is indicative of normalized intensity. The solid horizontal lines marked $\Delta\lambda_g$ and $\Delta\lambda_w$ are the average spectral shifts of a microlaser excited on glass and water. The histograms show the full distribution of the spectral shifts. The dashed and dotted lines are the average value and standard deviation of the shifts, respectively. b) Sketch of the relative position of the micromirror and microlaser for each position (i)-(v).The size of the microlaser is drawn to scale.} 
\label{fig:fig5}
\end{figure}

Panel (i) of Figure 5a represents the micromirror at the lowest position, almost at the same level as the microlaser. Hence, here the lasing modes circulating in a nearly horizontal plane probed mainly water, despite the microlaser touching the glass. Correspondingly, the shift is closer to the water reference line than to the glass one. As the micromirror was translated to higher positions (panels (ii) to (iv)), the lasing plane tilted more and more towards the glass, and the shift from the reference point decreased. In panel (v), the micromirror and microlaser are placed 57 \textmu m above the glass substrate. As expected, most modes show a clear shift away from the glass condition and are placed much closer to the water reference. In each panel, particularly in panel (iv), occasionally, the shifts depart from the reference value obtained with the objective. However, it can be seen that the most significant departure from the reference occurs for lasing peaks with low SNR (small radius of the symbol). For the sake of clarity, we did not take into account a few noise-generated nonphysical outliers with very low SNR and shift superior to four times the median of the peak fitting error values at a given micromirror position (see e.g., the outlier in panel (i)). The data associated with this paper are openly available and contain the complete set of spectra.  

\begin{figure}[ht!]
\centering\includegraphics[width=16cm]{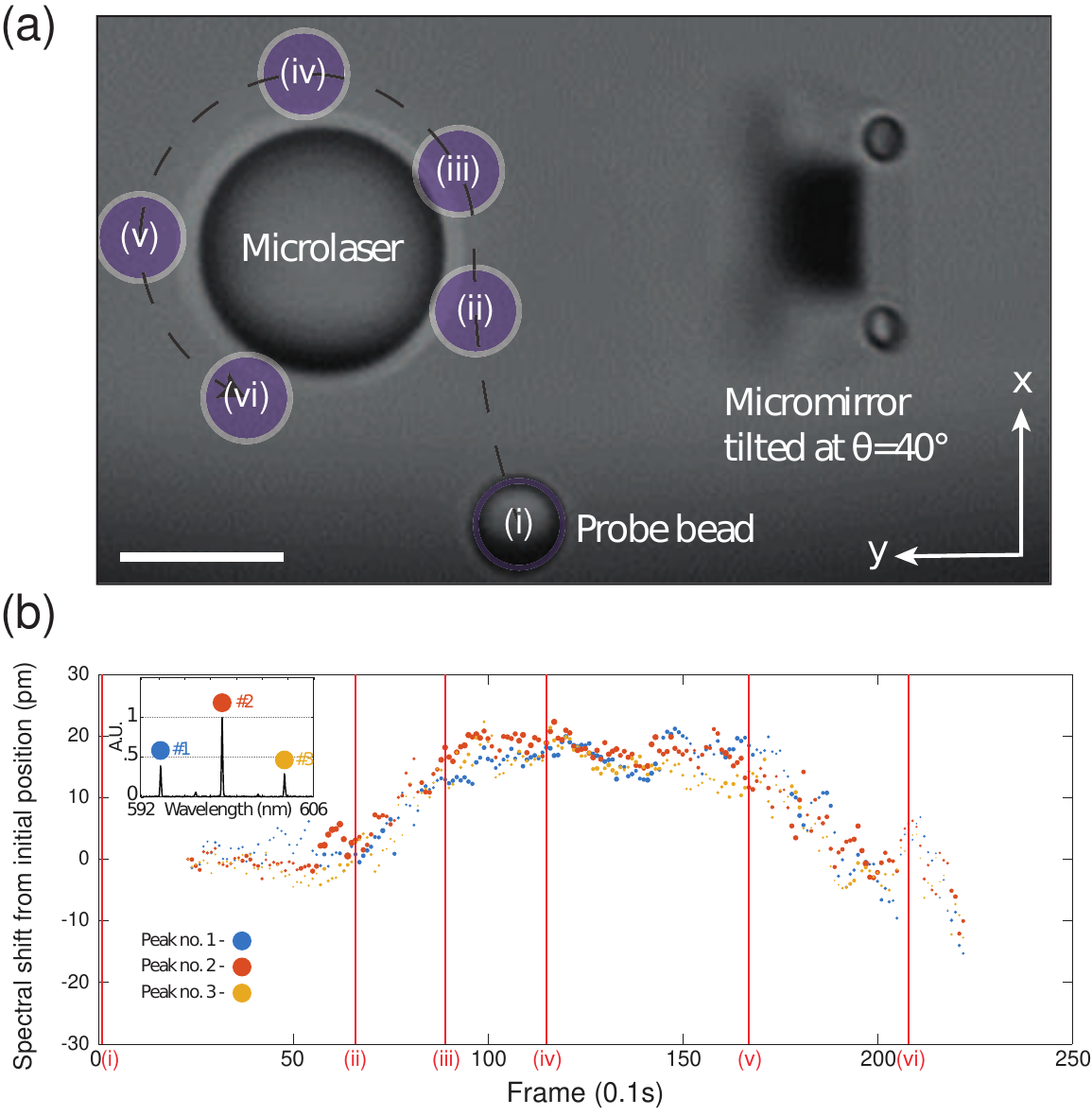}
\caption{a) Geometry for probing localized perturbations of refractive
index. The micromirror was used to illuminate a trapped microlaser while a polystyrene probe bead (diameter 5 \textmu m) was used to locally shift the refractive index at frame timestamps (i) 1, (ii) 66, (iii) 89, (iv) 115, (v) 167, and (vi) 208 b) Corresponding time trace of spectral shifts of three selected WGM peaks, showing displacement of peaks relative to their initial position in frame 20. The vertical red lines indicate the spectra at the position shown in (a). The initial normalized spectrum is shown in the inset. As before, the data points' size corresponds to the spectra' normalized intensity. The scale bar is 10 \textmu m. } 
\label{fig:fig6}
\end{figure}

\subsection{Probing localized perturbations of refractive index}
To demonstrate the versatility and unique capability of the micromirror platform, we used it to monitor a local dynamic change of the refractive index in the vicinity of the WGM microlaser. We simultaneously optically manipulated the micromirror and a 5 \textmu m polystyrene probe bead with refractive index $n_b = 1.59$, which we actively moved around the trapped and suspended microlaser, while acquiring the emission spectra in real-time. The micromirror, microlaser and probe bead were trapped 50 \textmu m above the glass substrate in order to remove any surface effects, with the centre of all objects located in the imaging plane of the microscope objective. \textbf{Figure 6a} shows a still image of the three objects and the path taken by the probe bead between positions marked (i) - (vi).

Figure 6b shows a typical time trace of the shift of the lasing modes that contains two hundred spectra acquired over 20 s, synchronized with a video acquired with the CCD (shown in SV1, Supplementary Information). In this case, to maintain a high SNR, we monitored the shift of only three most intense peaks for each spectrum (see inset in Fig. 6b for illustration). The shift was referenced to the emission spectrum obtained at frame 20 of the sequence, when the pump laser was switched on but before the probe bead touched the microlaser.

In the video SV1, Supplementary Information and Figure 6b, the pump laser was switched on at around frame 20. The bead was then slowly moved towards the microlaser, first making contact at frame 66 [position marked (ii)], which initiates a significant shift in the position of the peaks. The observed red-shift is consistent with what is expected from the presence of a high-index probe in the vicinity of the microlaser. The bead continues to orbit the microlaser around the equatorial plane before shifting up along the z-axis (out of the plane of the image) at frame 167 [position marked (v)], resulting in a gradual spectral shift back down to the reference spectra [position marked (vi)]. It should be noted that the probe bead does not follow an equatorial path but slightly changes height as it orbits the microlaser. This behavior is expected for optically trapped interacting beads, as due to the transverse oscillation of the larger particle, through hydrodynamic coupling, it will act on the smaller one\cite{LI2008135,PhysRevE.81.051403}. It is, therefore, not trivial to infer information about the exact position of the bead around the equatorial plane of the microlaser from the spectral shift.

\section{Conclusion}

Here, we demonstrate the application of microscopic optically trapped mirrors to precisely guide a nanosecond pump laser beam onto an active lasing microcavity formed by a dye-doped
polystyrene microsphere. The design, dimensions, and mechanical properties of the micromirrors allow accurate control over the angle, height, and distance of the pump beam with respect to the microlaser. This unprecedented control over the laser operation enables spatially resolved probing of the refractive index of the environment surrounding the microlaser. This was demonstrated by clearly exciting and distinguishing modes oscillating along planes in contact with the glass or only with the water.

The key constraint of the platform is the ability to orient the micromirror at angles which place the manipulated handles at positions away from the focal plane of the microscope objective. The low depth of field of the objective means the the Fresnel lenses generated by the SLM causes aberrations in the trapping beam, causing it to become unable to trap past a certain point. Employing lower NA objectives would address this issue and would also increase the field of view at the expense of the trap stiffness. Another solution would be to place the micromirror into a pre-programmed position, using either software approaches\cite{Eriksen2002FullyDM} or using holographic traps encoded on microfluidic chips\cite{doi:10.1021/acsphotonics.2c01986}. A further constraint is given by geometrical considerations. The mirror in the center of the membrane acts as an aperture for the focused pump beam. Therefore, there exists a trade off between the distance of the mirror from the microlaser and the spot size of the pump beam and the mirror size.

The intrinsic scattering of biological samples mitigates the reduced collection efficiency of the modes lasing on a plane perpendicular to the collection plane. This could be further increased using a second micromirror to collect and redirect the signal back to the objective. Additionally, the pumping efficiency could be improved by replacing the gold mirror with a dielectric one, thus allowing more light to reach the microlaser.

The platform also presents an opportunity of extending the capabilities of the micromirrors through textured metasurfaces, e.g., to manipulate the wavefront and momentum of the light incident on the membranes\cite{Andren2020MicroscopicMP}, and take advantage of the intrinsic stability of trapped photonic membranes\cite{Askari:21} for advanced interaction with biological specimens. These advanced micromirrors can be used to generate complex illumination light fields akin to light sheets, that can be scanned across e.g. a cell. This approach will enable the real-time tomography of the specimen, using a single objective.

To conclude, we presented a novel and versatile platform uniquely suited to interact with extended capability and degrees of freedom with microscopic objects in microfluidic environments. We demonstrated the ability to pump WGM lasers on planes perpendicular to the axis of the microscope objective. The technique can be used in real-time to monitor the spatially resolved change of refractive index of the environment surrounding the microlasers. Our platform contributes to the ongoing effort to bring photonic micro tools into microfluidic environments, focusing on improving how we deliver and collect light from samples.


\section{Experimental Section}

\threesubsection{Trapping Experiments}

For trapping experiments, we placed a solution of microlasers (PS-FluoRed, 15.35 \textmu m particle diameter, Microparticles GmbH) and micromirrors diluted using D\textsubscript{2}O in a microfluidic chamber made by placing a 100 \textmu m thick vinyl spacer between a glass microscope slide and a 170 \textmu m thick coverslip. For the micromirrors, we had a concentration up to 500/100 \textmu l, calculated dissolving in a known amount of solution a known number of micromirrors. This is an approximation since we suspect that the yield of the process is not 100\%. For the microlasers, we diluted the concentration to a condition that guaranteed a sparse enough number of microlasers for experimental convenience (up to 1 microlaser visible in a field of view of 50x50 \textmu m). The typical lifetime of the solution would be 1-2 weeks, after which the liquid would begin to evaporate. The experimental protocol required first locating the micromirrors, trapping them, and placing them near a microlaser. The microlasers were optically trapped to ensure a consistent position using the same holographic tweezer setup.  

The micromirror center was placed 40 \textmu m away from the microlaser center along the y direction to ensure the pump beam would not excite the microlaser directly without first reflecting from the mirror. The distance was measured by calibrating the imaging camera with a reference sample. The pump beam was defocused by 17 \textmu m relative to the imaging plane of the microscope objective to ensure the reflective section of the micromirror was overfilled.
The measured average power of the pump laser before the objective was $\sim$0.7 mW. A ray diagram can be found in Figure S2, Supplementary Information. The width of the beam at the position of the microlaser surface was estimated using a Gaussian beam model to be $\sim$8 \textmu m, and we used a maximum average pump power of 2 mW (which would be reduced to 0.7 mW by the time the pump laser reached the objective) to reduce heating effects at the metallic mirror.

Empirically, the optimal position of the center of the micromirror was found to be 3.5 \textmu m away from the imaging plane of the objective in the direction of pump beam propagation, which allowed for sampling more of the imaging plane without the necessity for 45° tilts as shown in Figure S3, Supplementary Information. 

A typical lasing spectrum contained more than 10 modes. However, to monitor the shifts due to the refractive index change, we considered the average of the shift of the central six modes, to increase the signal-to-noise ratio. The averaging also allowed us to consider uncertainties due to the heating and motion of the microlaser. Each spectrum was acquired five times over 1 s during experiments with a stationary bead to increase the signal-to-noise ratio, and ten times over 1 s during the probing of localized perturbations to increase the temporal resolution. 

To precisely manipulate the micromirrors, a custom LabVIEW software utilizing a console controller was used to address the SLM. Blazed gratings and Fresnel lenses were projected and used to displace the beams 
in x, y, and z, respectively, allowing for up to 6 independent trapping sites by overlaying the patterns. Each trapping beam had an average power of 2 mW at the sample plane, and the trapping power was distributed equally among individual traps. The micromirror traps were connected together using the software. A normal vector from the micromirror was tracked to allow for full 6 degrees of motion using two buttons to go up and down in z, two buttons for rotating in the xy plane, one joystick for movement in the xy plane and one joystick for controlling the pitch and roll angles. A noise mask was also applied to the SLM to decrease ghost orders arising from the trapping array's high symmetry.\cite{jones_maragò_volpe_2015}

\medskip
\textbf{Supporting Information} \par 
Supporting Information is available from the Wiley Online Library.

\medskip
\textbf{Acknowledgements} \par 
We acknowledge Henry Archer's contribution to the development of the software used to manipulate the micromirrors. 
The project was supported by the European Research Council (ERC) under the European Union Horizon 2020 research and innovation program (Grant Agreement No. 819346). M.C.G. acknowledges financial support through an Alexander von Humboldt Professorship. 
\medskip \\
\textbf{Data availability} 
The data that support the findings of this study are openly available in the Data Set of the University of St. Andrews Research Portal at\\https://doi.org/10.17630/413a0461-d941-43db-bb7f-379dc05ec0bc.
\medskip \\
\textbf{Conflict of Interest}
The authors declare no conflicts of interest.

\newpage

%
\bibliographystyle{MSP}
\bibliography{Main}

\end{document}